\documentclass[aps,prb,superscriptaddress,twocolumn,showpacs]{revtex4}

\usepackage{graphicx}

\begin{document}

\title{On Order-Disorder (L$2_1$ $\to$ B$2^\prime$)
Phase Transition \\in Ni$_{2+x}$Mn$_{1-x}$Ga Heusler Alloys}

\author{V.~V.~Khovailo}
\affiliation{Institute of Fluid Science, Tohoku University, Sendai
980-8577, Japan}
\author{T.~Takagi}
\affiliation{Institute of Fluid Science, Tohoku University, Sendai
980-8577, Japan}
\author{A.~N.~Vasil'ev}
\affiliation{Physics Faculty, Moscow State University, Moscow
119899, Russia}
\author{H.~Miki}
\affiliation{Faculty of Systems, Science and Technology, Akita
Prefectural University, Honjo 015-0055, Japan}
\author{M.~Matsumoto}
\affiliation{Institute of Multidisciplinary Research for Advanced
Materials, Tohoku University, Sendai 980-8577, Japan}
\author{R.~Kainuma}
\affiliation{Graduate School of Engineering, Tohoku University,
Sendai 980-8577, Japan}

\pacs{64.60.Cn, S1.2}

\maketitle

The ferromagnetic $\textrm{Ni}_{2+x}\textrm{Mn}_{1-x}\textrm{Ga}$
Heusler alloys are of considerable interest due to their potential
applicability as magnetically driven shape memory materials
\cite{Vas1}. For the stoichiometric composition of
$\textrm{Ni}_2\textrm{MnGa}$ the melting temperature is 1382~K
(Ref.~\onlinecite{Over}). At cooling from the liquid phase these
triple alloys usually solidify in the disordered A2 phase
characterized by an arbitrary occupation of every site in the
crystal lattice. In principle, the chemical ordering in solid
state of Heusler alloys is possible either through an intermediate
partially ordered B$2^{\prime}$ phase or directly to the
completely ordered body-centered cubic L$2_1$ phase \cite{McC}. In
the B$2^{\prime}$ phase Ni atoms order while Mn and Ga atoms
occupy their sites in the crystal lattice randomly. However, the
neutron diffraction measurements as well as the differential
thermal analysis \cite{Over} of the Ni-Mn-Ga system presented no
clear evidence for the presence of the $\textrm{A}2 -
\textrm{B}2^{\prime}$ phase transition, meaning presumably that
these alloys solidify in the partially ordered B$2^{\prime}$
phase. For the stoichiometric composition of
$\textrm{Ni}_2\textrm{MnGa}$ the $\textrm{L}2_1 \to
\textrm{B}2^{\prime}$ phase transition temperature is 1071~K.

At further cooling, the
$\textrm{Ni}_{2+x}\textrm{Mn}_{1-x}\textrm{Ga}$ Heusler alloys
undergo a structural phase transition from the body-centered cubic
phase to the body-centered tetragonal ($c/a = 0.94$) martensitic
phase, characterized by the pronounced effects of shape memory and
superelasticity \cite{Ull,OH,Tickle}. For the stoichiometric
$\textrm{Ni}_2\textrm{MnGa}$ composition the martensitic
transition temperature is equal to 202~K. This transition is
preceded by a premartensitic transition at 260~K, which is the
formation of the static displacement waves in the lattice with the
wavevector $[\frac{1}{3} \frac{1}{3} 0]$
(Refs.~\onlinecite{Zhel,Stu}). The deviation from stoichiometry in
$\textrm{Ni}_{2+x}\textrm{Mn}_{1-x}\textrm{Ga}$ alloys results in
merging of the premartensitic and martensitic transitions, so that
the tetragonal phase appears to be modulated by the static
displacement waves.

While being in cubic L$2_1$ phase, $\textrm{Ni}_2\textrm{MnGa}$
exhibits a ferromagnetic phase transition with the Curie
temperature $T_C = 376$~K. The change of composition in the
$\textrm{Ni}_{2+x}\textrm{Mn}_{1-x}\textrm{Ga}$ system results in
a decrease of Curie temperature and an increase of martensitic
transition temperature until they merge at $x = 0.18 - 0.20$
(Ref.~\onlinecite{Vas2}). While the low temperature phase
transitions in the $\textrm{Ni}_{2+x}\textrm{Mn}_{1-x}\textrm{Ga}$
system were studied in many aspects, the high temperature
$\textrm{L}2_1 \to \textrm{B}2^{\prime}$ phase transition needs
further investigation. In the present work we studied this
transition in $\textrm{Ni}_{2+x}\textrm{Mn}_{1-x}\textrm{Ga}$ $(x
= 0.16 - 0.20)$ by means of differential scanning calorimetry
(DSC) measurements.

The $\textrm{Ni}_{2+x}\textrm{Mn}_{1-x}\textrm{Ga}$
polycrystalline ingots were prepared by a conventional arc-melting
method under argon atmosphere. The ingots were annealed at 1100~K
for nine days in quartz ampoules and quenched in ice water.
Samples for the measurements were cut from the middle part of the
ingots. The measurements were done by a NETZSCH-404 high
temperature differential scanning calorimeter in the temperature
range from 750 to 1200~K.

\begin{figure}[h]
\begin{centering}
\includegraphics[width=\columnwidth]{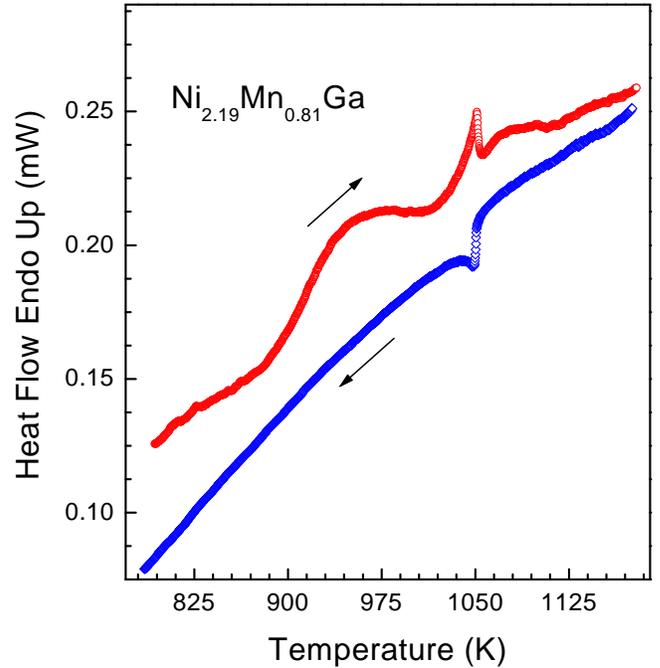}
\caption{DSC curves measured at heating and cooling for
Ni$_{2.19}$Mn$_{0.81}$Ga.}
\end{centering}
\end{figure}

The results of DSC measurements for the
$\textrm{Ni}_{2.19}\textrm{Mn}_{0.81}\textrm{Ga}$ sample are shown
in Fig.~1. The well-defined peaks on the curves obtained upon
heating and cooling correspond to the $\textrm{L}2_1 \to
\textrm{B}2^{\prime}$ phase transition. It is evident that these
anomalies are of a characteristic $\lambda$-type expected for a
second order transition. No difference in the temperatures of this
phase transition in the heating and cooling cycles was detected
within the experimental uncertainties. In fact, the same
$\lambda$-type anomalies were seen also at $\textrm{B}2 -
\textrm{L}2_1$ second order phase transition in the Cu-Al-Mn
system \cite{Obrado}.

\begin{figure}[t]
\begin{center}
\includegraphics[width=\columnwidth]{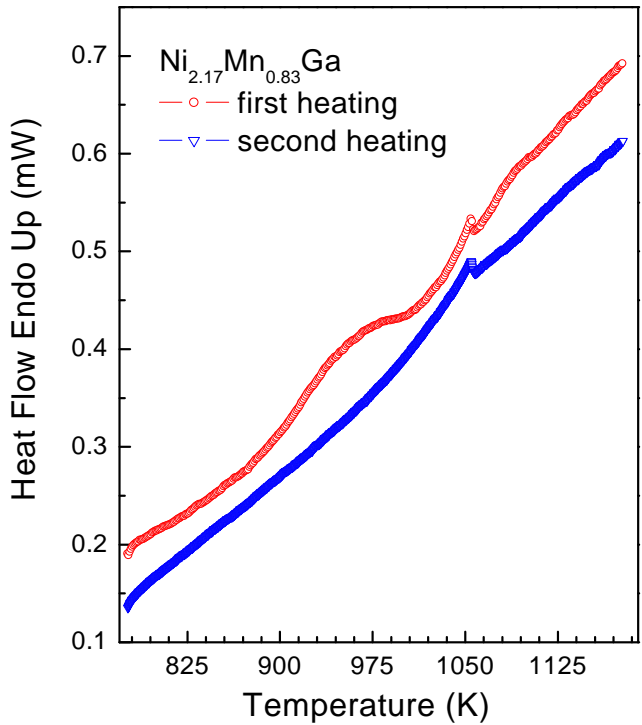}
\caption{Heat flow as a function of temperature during first and
second heating in Ni$_{2.17}$Mn$_{0.83}$Ga.}
\end{center}
\end{figure}

An additional feature of these measurements is the appearance of a
broad anomaly in a temperature range from 870 to 1030~K on the
heating curve. The position of this anomaly varied randomly below
the temperature of the $\textrm{L}2_1 \to \textrm{B}2^{\prime}$
second order phase transition, but it was seen in every sample
studied. To clarify the origin of this extra contribution, the
$\textrm{Ni}_{2.17}\textrm{Mn}_{0.83}\textrm{Ga}$ samples were
subjected to repeated heating and it was found that the broad
anomalies vanish just after the first heating - cooling cycle. The
results of DSC measurements obtained during first and second
heating process for the
$\textrm{Ni}_{2.17}\textrm{Mn}_{0.83}\textrm{Ga}$ sample are shown
in Fig.~2. Based on these measurements it can be concluded that
the additional anomalies on DSC curves appear due to the procedure
of the thermal treatment. The quenching of the samples resulted in
the stabilization of the partially ordered B$2^{\prime}$ phase. At
first heating randomly distributed Mn and Ga atoms occupy their
sites in the L$2_1$ structure and this process is accompanied by
an additional heat exchange with the calorimeter. At repeated
heating these broad anomalies disappear since in the first heating
and cooling cycle the samples were slowly cooled from the
disordered state. Assumingly, during this procedure the Mn and Ga
atoms took their sites in the L$2_1$ structure.

\begin{figure}[t]
\begin{center}
\includegraphics[width=\columnwidth]{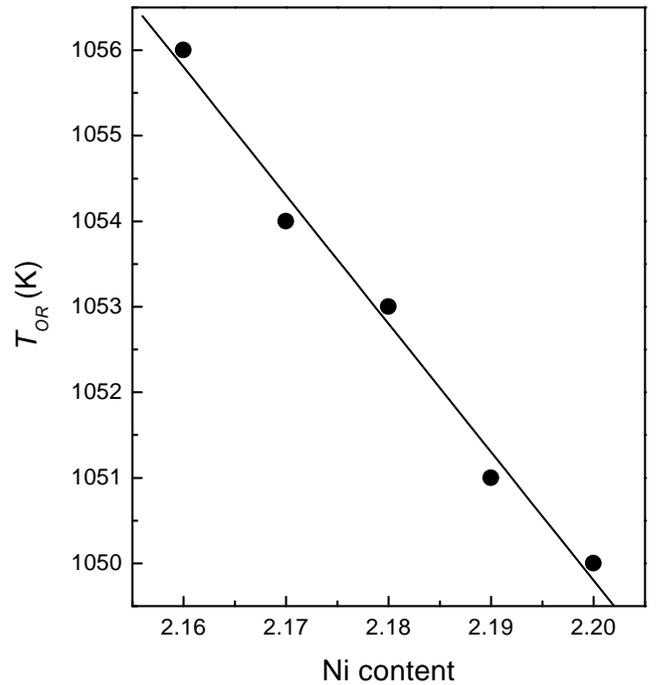}
\caption{The compositional dependence of the $\textrm{L}2_1 \to
\textrm{B}2^\prime$ phase transition temperature in
$\textrm{Ni}_{2+x}\textrm{Mn}_{1-x}\textrm{Ga}$ ($x = 0.16 -
0.20$).}
\end{center}
\end{figure}

The compositional dependence of the chemical ordering temperature
$T_{OR}$ is shown in Fig.~3. The $\textrm{L}2_1 \to
\textrm{B}2^{\prime}$ phase transition temperature somewhat
decreases with the increase of the Ni content in
$\textrm{Ni}_{2+x}\textrm{Mn}_{1-x}\textrm{Ga}$ alloys. In fact,
this tendency could be probably expected since the $\textrm{L}2_1
\to {\textrm B}2^{\prime}$ phase transition temperature in
$\textrm{Ni}_3\textrm{Ga}$ is much lower than in
$\textrm{Ni}_2\textrm{MnGa}$ $(x = 0)$, and the former compound
could be considered formally as a limiting case of a triple alloy
composition with $x = 1$ (Ref.~\onlinecite{Bin}).

In conclusion, it was found that the $\textrm{L}2_1 \to
\textrm{B}2^{\prime}$ phase transition in
$\textrm{Ni}_{2+x}\textrm{Mn}_{1-x}\textrm{Ga}$ $(x = 0.16 -
0.20)$ Heusler alloys is of second order and the temperature of
this transition decreases with Ni excess. Extra anomalies observed
in DSC measurements during the first heating - cooling cycle below
the $\textrm{L}2_1 \to \textrm{B}2^{\prime}$ transition appear
presumably due to the ordering of non-equilibrium B$2^{\prime}$
phase obtained through quenching of the samples from high
temperatures to the equilibrium L$2_1$ phase.

\subsection*{Acknowledgements}
This work was partially supported by the Grant-in-Aid for
Scientific Research (C) No.~11695038 from the Japan Society of the
Promotion of Science and by Grant-in-Aid of the Russian Foundation
for Basic Research No.~99-02-18247.

\end{document}